# Superconductivity in a ferroelectric-like topological semimetal SrAuBi

Hidefumi Takahashi[1,2*], Tomohiro Sasaki[1], Akitoshi Nakano[3], Kazuto Akiba[4], Masayuki Takahashi[1], Alex H. Mayo[1,5], Masaho Onose[1,6], Tatsuo C. Kobayashi[4], and Shintaro Ishiwata[1,2†]

[1]*Division of Materials Physics and Center for Spintronics Research Network (CSRN), Graduate School of Engineering Science, Osaka University, Osaka 560-8531, Japan*

[2]*Spintronics Research Network Division, Institute for Open and Transdisciplinary Research Initiatives, Osaka University, Yamadaoka 2-1, Suita, Osaka, 565-0871, Japan*

[3]*Department of Physics, Nagoya University, Nagoya 464-8602, Japan*

[4]*Graduate School of Natural Science and Technology, Okayama University, Okayama 700-8530, Japan*

[5]*Institute for Materials Research, Tohoku University, Sendai 980-8577, Japan*

[6]*Department of Applied Physics, The University of Tokyo, Tokyo 113-8656, Japan*

Given the rarity of metallic systems that exhibit ferroelectric-like transitions, it is apparently challenging to find a system that simultaneously possesses superconductivity and ferroelectric-like structural instability. Here, we report the observation of superconductivity at 2.4 K in a layered semimetal SrAuBi characterized by strong spin-orbit coupling (SOC) and ferroelectric-like lattice distortion. Single crystals of SrAuBi have been successfully synthesized and found to show a polar-nonpolar structure transition at 214 K, which is associated with the buckling of Au-Bi honeycomb lattice. On the basis of the band calculations considering SOC, we found significant Rashba-type spin splitting and symmetry-protected multiple Dirac points near the Fermi level. We believe that this discovery opens up new possibilities of pursuing exotic superconducting states associated with the semimetallic band structure without space inversion symmetry and the topological surface state with the strong SOC.

*takahashi.hidefumi.es@osaka-u.ac.jp
†ishiwata.shintaro.es@osaka-u.ac.jp

**INTRODUCTION**

Metallic materials have been considered to unfavor ferroelectric-like structure transitions as the induced ferroelectric dipole tends to be screened by itinerant electrons[1,2]. Despite this tendency, ferroelectric-like structural transitions have recently been found with great attention in several metallic compounds, such as $LiOsO_3$ and $β$-$MoTe_2$[3–5]. As is evident from the scarcity of ferroelectric-like metals, superconductors that exhibit ferroelectric-like structural transition are even rarer, since superconductivity typically favors metallic states with large amounts of itinerant electrons.[6–10]. Nevertheless, the coexistence of the superconductivity and the ferroelectric-like transition has been actively discussed, as they have a potential to exhibit unique features typified by the singlet-triplet mixing state[11,12] and topological superconductivity with a Majorana edge state[13,14]. For instance, the discovery of a polar superconductor $CePt_3Si$ has spurred theoretical and experimental investigations inherent in the heavy fermion systems with inversion symmetry breaking[12,15–18].

One of the most important factors yielding such unique electronic properties in these superconductors is the spin-orbit coupling (SOC), which is especially important in noncentrosymmetric systems containing heavy elements such as Bi. For instance, half-Heusler compounds R(Pt, Pd)Bi (R is a rare-earth metal) with a noncentrosymmetric structure and strong SOC have been considered to be unconventional superconductors with quintet or septet Cooper pairing dominated by the $j=3/2$ electrons ($j$ is the total angular momentum)[19–22]. To be noted here is that the role of SOC in polar structures differs from that in noncentrosymmetric and nonpolar structures in that it produces electromagnetic effects and nonreciprocal responses[23–26]. The effect of SOC on superconductivity manifests itself especially in topological semimetals, where surface superconductivity derived from nontrivial surface states has been reported[27–34]. Among them, the surface state (Fermi arc) of Weyl semimetals without inversion symmetry (but with time-reversal symmetry) is closely related to the bulk state; i.e., the penetration depth of the surface states into the bulk depends strongly on the surface momentum, and the open strings of the surface states connect the bulk Bloch waves at the Weyl nodes of opposite chirality[35]. This situation raises the question of whether the surface superconductivity can be realized independently of the bulk state. Therefore, topological semimetals without inversion symmetry provide fundamental investigations for 2D superconductivity and innovative interface functionality.

Here, we discover superconductivity as well as a ferroelectric-like structural transition in a topological semimetal SrAuBi. This compound crystallizes in the hexagonal ABC-type structure, where the Sr and Au-Bi atoms form two-dimensional triangular lattice and honeycomb lattice, respectively (Fig 1a)[36]. While SrAuBi is located at the phase boundary between the polar and nonpolar structures in the phase diagram of the ABC-type compounds[37], it has been reported to be a centrosymmetric system and the detailed structure has not been clarified. In this study, we obtained single crystals of SrAuBi and found a ferroelectric-like polar-nonpolar structure transition at 214 K and superconductivity at 2.4 K in the polar structure. The band calculations suggest the Rashba-type spin splitting and the presence of symmetry-protected Dirac points near the Fermi level, possibly inducing the unconventional superconductivity associated with the surface state.

**RESULTS**

**Crystal structure**

Figure 1(c) shows the XRD pattern of SrAuBi (space group: $P6_3/mmc$) with a small amount of the Bi flux (Elemental analysis of a single crystal by SEM/EDX is shown in Supplementary Note 1). In order to refine the crystal structure of SrAuBi at selected temperatures, we performed single-crystal XRD measurements as shown in Figs. 2(a) and 2(b). The XRD patterns are shown in Supplementary Note 2. Whereas the XRD profile at room temperature can be indexed by the hexagonal unit cell with space group $P6_3/mmc$ ($a = b = 4.7915(7)$ Å, $c = 8.6737(7)$ Å), that below 214 K is well refined by the polar and hexagonal structure with space group $P6_3mc$ ($a = b = 4.7822(2)$ Å, $c = 8.6402(2)$ Å at 110 K), which is isostructural with EuAuBi[38]. Detailed analysis to verify the inversion symmetry breaking upon the polar lattice distortion is described in Supplementary Note 3. Corresponding to the structural phase transition, an anomaly is observed in the temperature dependence of the $c$-axis length as shown in Fig 2(c). This structural phase transition is characterized by the buckling of the

Au-Bi honeycomb lattice, which produces polarity along the *c* axis through the staggered shift in the coordinates of *z* of Au and Bi sites; $Au^+$ and $Bi^{3-}$ ions alternately shift in the *c*-axis direction. From these experimental results, we have revealed for the first time that SrAuBi exhibits a ferroelectric-like polar-nonpolar structure transition at 214 K. In order to examine the stability of polar ($P6_3mc$) and nonpolar ($P6_3/mmc$) structures for SrAuBi, the enthalpies *H* were calculated based on *ab*-initio calculations. Notably, the polar phase ($H_p$ = -52841.954 eV) is more stable than the nonpolar phase ($H_{np}$ = -52841.935 eV), and the energy scale of the enthalpy difference ($\Delta H = H_p - H_{np}$ = -0.019 eV) is comparable to the structural transition temperature. This result is consistent with the fact that SrAuBi is located at the boundary between polar and nonpolar phases in the structural phase diagram of the hexagonal ABC-type compound.[37]

**Superconducting properties**

The temperature dependence of the in-plane (*I*//*ab*) electrical resistivity ρ is shown in Fig. 3(a). Metallic behavior is observed below 300 K. A weak anomaly in the temperature dependence of electrical resistivity is observed around 210 K, which is clearly seen in the temperature derivative for the resistivity *d*ρ/*dT*, corresponding to the structural phase transition. It is noteworthy that the resistivity starts to drop near 2.5 K ($T_c$ onset) and reaches zero below 2.4 K ($T_c$ zero), indicating the emergence of superconductivity, just like the cases for EuAuBi[38]. To evaluate the volume fraction of superconductivity, we measured the magnetic susceptibility (4πχ) with the out-of-plane field *H* = 0.008 mT (*H*//*c*), as shown in Fig. 3(b). A Meissner effect is observed below 2.4 K, and the superconducting volume fraction reaches about 20 % (for the measurements of a Meissner effect with demagnetization correction and in ZFC and FC processes, see Supplementary Notes 5 and 6). Since the lowest measured temperature 1.8 K is so close to $T_c$ that the estimated volume fraction is expected to exceed 20 %. Furthermore, the Meissner effect disappears by the application of rather low magnetic fields (> 0.1 mT), implying that magnetic flux penetrates in weak magnetic fields as in the cases of noncentrosymmetric superconductors YPtBi and EuAuBi[38,39], which presumably reflects the smallness of the electronic density of states at Fermi level inherent in the semimetallic systems.

The temperature dependence of the in-plane resistivity ρ measured in the out-of-plane (*H*//*c*) and the in-plane (*H*//*ab*) magnetic fields up to 1.5 T above 1.8 K is shown in Fig. 4(a) and (b) for the samples No. 1 and 2. The insets of Fig. 4(a) and 4(b) show the results down to 0.3 K with different samples No. 3 and 4. The critical temperature $T_c$ decreased upon the application of the magnetic fields, which supports the emergence of a superconducting phase. The critical temperatures are defined by the onset of resistivity drop ($T_c$ onset) and by zero resistivity ($T_c$ zero). In addition, shoulder structures are identified in ρ between $T_c$ onset and $T_c$ zero especially for *H*//*c*, when the magnetic field is larger than 2 T. This anomalous behavior suggests the inhomogeneity of superconducting states associated with the structural domain, surface state, and decomposition of crystal surfaces. In the time-dependent powder X-ray diffraction patterns for SrAuBi, the peaks of $Au_2Bi$ and Bi develop gradually over time, which can be associated with the decomposition of crystal surfaces (Supplementary Fig. 8).

To gain more insight into the superconducting state, we conducted the magnetic field dependence of in-plane $\rho$ with $H//c$ and $H//ab$ for samples 3 and 4, as shown in Supplementary Fig. 9 (Supplementary Note 7). $\rho(H)$ exhibited shoulder-like unusual structures below 1 K, as observed in $\rho(T)$. The superconducting state turned to the normal state at ~ 5 T and 4.1 T for $H//c$ and $H//ab$, respectively. The critical fields are defined by the onset of resistivity drop ($H_c$ onset) and by zero resistivity ($H_c$ zero).

By summarizing the resistivity data, we established the superconducting phase diagrams of SrAuBi as functions of $H$ and $T$ for $H//c$ and $H//ab$, as shown in Figs. 4(c) and 4(d), respectively. The temperature dependences of critical fields $H_c$ have a convex downward near $T_c$, characteristic of the multiband/multigap superconductors such as $MgB_2$[40,41], SrPtAs[42], RPdBi[43], and Fe-based superconductors[44,45]. The $H_c$ with zero resistivity and onset of superconducting transition at the lowest temperature for $H//c$ is ~ 3.0 T and 5.0 T, respectively. For $H//ab$, the $H_c$ values are small [~ 2.8 T ($H_c$ zero) and ~4.1 T ($H_c$ onset)] and weakly anisotropic. These values are much larger than that of a similar hexagonal material of SrPtAs ($H_c$ ~0.22 T)[42] and are comparable or slightly larger than the Pauli limited upper critical field $H_P$ of ~4.3 T evaluated from the equations $H_P = \Delta/\sqrt{2}\mu_B$ and $\Delta = 1.76k_BT_c$. The observation of such high $H_c$ values precluded the possibility of superconductivity originating from impurities, because such unusually high $H_c$ values have not been reported in alloys and nanoparticles of Bi or $Au_2Bi$[46–48]. In addition, the high and anisotropic $H_c$ is also observed in EuAuBi [$H_c$ ($H//c$)~ 10 T and [$H_c$ ($H//ab$)~ 3 T ], presumably reflecting the Rashba-type spin splitting associated with the polar structure, crystal surface, and strong SOC; the larger value and significant anisotropy of $H_c$ for EuAuBi may stem from the large magnetic moment of $Eu^{2+}$ ($S=7/2$) which enhances spin splitting[38].

For bulk superconductivity, a specific heat jump is expected. However, it turns out that a jump in specific heat is absent, as shown in Supplementary Fig. 10 (Supplementary Note 8). There are two possible explanations for this result. The first is that although the material is bulk superconducting, the specific heat jump is inherently weak reflecting the small electronic specific heat coefficient (~ 0.8 mJ/mol $K^2$), as discussed in YPtBi[49]. In addition, the magnetic flux seems to penetrate in weak magnetic fields such as geomagnetism, suppressing the specific heat jump.

Another possibility is superconductivity in the surface state derived from the topological band structure as reported for topological semimetals[30,32]. We examined the sample-size dependence of the superconducting critical current to see whether the superconductivity in SrAuBi is emerging as a bulk or a surface state. We measured the electrical resistivity of one sample by varying the sample thickness $D$ to $D = 0.56$ mm (No. S1), 0.44 mm (No. S2), 0.27 mm (No. S3), and 0.16 mm (No. S4). Supplementary Figure 11 shows the current dependence on the resistivity under these conditions. Even if the thickness $D$ is changed, the superconducting transition appears at almost the same temperature when the applied current value is the same. This result rules out the possibility of bulk superconductivity, since $T_c$ of bulk superconductivity should be suppressed by the increment of the current density, which increases with decreasing $D$. On the other hand, the result is consistent with the surface superconductivity, which is considered insensitive to the sample thickness and the current density. Figure 5 displays the superconducting transition temperature versus current density and current value for samples with different thicknesses. The $T_c$ is scaled not with the current density but with the current values, suggesting the emergence of the surface superconductivity, as discussed for BaMg$_2$Bi$_2$.[32] On the other hand, this unusual cross-section-dependence of $I_c$ rules out the filamentary superconductivity because $I_c$ should depend on the cross-section (volume) of the sample, when considering filamentary superconductivity randomly distributed in the crystal.

At present, it is not clear whether superconductivity in SrAuBi is of bulk or surface origin. Assuming that the bulk superconductivity is the case, the unusual superconducting state proposed for the Y(Pd,Pt)Bi system can be considered[22]. On the other hand, while a relatively large Meissner effect is observed as a signature of bulk superconductivity, we obtained other experimental results supporting the surface superconductivity, such as the absence of superconducting jump in specific heat and the significant sample-thickness dependence of the superconducting critical-current density. In particular, topological semimetals potentially exhibit the coexistence of bulk and surface superconductivity[35]. In the following, we discuss the characteristic electronic states and superconducting properties based on band calculations, leaving open the possibility of both bulk and surface superconductivity.

**DISCUSSION**

To examine the effect of the polar lattice distortion on the band structures and superconducting properties, we performed first-principles band calculations with and without SOC. As shown in Fig. 6(a), the semimetallic band structure can be confirmed by the presence of small hole pockets around the Γ point and small electron pockets at M and A points. Reflecting the semimetallic band structure, density of states near the Fermi level is found to be extremely small [Fig. 6(a)]. One of the key features is the occurrence of a band inversion around the Γ-point owing to the conduction bands of the Bi 6$p$ orbital and electron bands of the Au 6$s$ orbital, which is also similar to the half-Heusler superconductors[22]. As a result, a gapless Dirac dispersion appears slightly above $E_F$ along the Γ–M direction, which corresponds to the in-plane direction ($ab$-plane), when SOC is not taken into account [solid black line in Figs. 6(a) and 6(b)]. When SOC is included (solid red line), the band crossing gaps out, resulting in the formation of a massive Dirac dispersion [Fig. 6(b)]. In addition, the bands slightly above Fermi energy around Γ point (Γ–M direction) have the finite Rashba-type spin-splitting inherent to the polar structure; the bands described in red lines split along the Γ–M direction. Given such a Rashba-type splitting, the Pauli limiting field ($H_c$) is strongly enhanced by the parity-breaking SOC for the out-of-plane field[17,50–53]. It is noted that the previous study for the same system predicts the emergence of Weyl points on the $k_z$=0 plane and off the high-symmetry lines[54]. Another key feature can be seen along the Γ–A line as shown in Fig. 6(c), which is the inter-layer direction parallel to the $c$-axis. There exists one Dirac-like point below $E_F$ and multiple Dirac-like points near $E_F$, when excluding and including SOC, respectively. The Dirac nodes on the Γ–A line are immune to SOC because of the protection by the sixfold rotational symmetry around the $c$ axis[54]. These topological band structures suggest the existence of a characteristic surface state, potentially responsible for surface superconductivity.

Finally, we discuss the possibility of nontrivial superconducting states in SrAuBi. Owing to the ferroelectric-like lattice distortion, the possibility of exotic superconducting states such as mixed singlet-triplet pairing can be considered for SrAuBi, as proposed for heavy-fermion superconductors. Based on band calculations, the relatively large $H_c$ is attributable to the significant Rashba-type spin splitting, which is one of the necessary conditions to achieve nontrivial superconducting states[55,56]. In addition, topological band structure, including the multiple Dirac points on Γ–A line and possible Weyl points on the $k_z$=0 plane, may influence the superconducting properties[13,57,58]. Such a topological electronic state can be a source of unique superconducting states emerging on the crystalline surface or the polar domain boundaries[28,48,59], which may coexist with the bulk superconductivity. In particular, SrAuBi has a semimetallic band structure with a small density of states at the Fermi level, which potentially induces superconductivity on the sample surface with Fermi arcs caused by the crystal inversion symmetry breaking, which is more stable than bulk superconductivity[35].

In conclusion, single crystals of the Bi-based compound SrAuBi were successfully synthesized by the self-flux method and found to show potentially unique superconductivity. The synchrotron X-ray diffraction measurements and resistivity measurements have revealed that SrAuBi exhibits a ferroelectric-like polar-nonpolar structure transition at 214 K, followed by superconductivity at 2.4 K. Point is that the upper critical magnetic field was estimated to be more than 4 T and was larger for the out-of-plane field (parallel to the polar axis) than that for the in-plane field as expected for the polar superconducting materials[15,50,52,53]. Considering the band calculations showing the Rashba-type spin-splitting and symmetry-protected multiple Dirac points, SrAuBi can be the prototypical and fertile material that engenders exotic superconducting properties.

## METHODS

### Sample preparation

Single crystals of SrAuBi [Fig. 1(b)] were grown using a Bi self-flux method. High-purity ingots of Sr (99.9%), Au (99.99%), and Bi (99.999%) were mixed at a ratio of Sr:Au:Bi = 1:1:10 and placed in an alumina crucible in an argon-filled glove box[60]. The crucible was sealed in an evacuated quartz tube and heated at 1100 °C for 10 h, followed by slow cooling to 400 °C at a rate of ~ 4 °C/h. The excess flux was decanted using a centrifuge.

### Sample Characterization

The powder X-ray diffraction (XRD) measurements were performed by a Bruker D8 advance diffractometer with Cu Kα radiation at room temperature. We conducted synchrotron X-ray diffraction measurements for single crystalline samples at BL02B1 in SPring-8. We used a Pilatus3 X 1M CdTe detector[61] for obtaining high-resolution two-dimensional diffraction patterns with a wavelength of 0.3100 Å. We crushed a bulk-size single crystal into micro-size, and picked up a small piece with a dimension of 80 ×50×25μm$^3$. Diffraction intensity averaging was performed with *SORTAV* [62], and crystal structure refinement was carried out by means of the *SHELXL* least squares program[63]. The chemical composition of the single crystals were analyzed by a scanning electron microscope (SEM) equipped with an energy-dispersive X-ray (EDX) analysis probe (Hitachi High-Tech, SU-9000).

**Magnetic and transport measurements**

The magnetic and transport properties above 1.8 K were measured using a magnetic property measurement system (MPMS, Quantum Design, Inc.). The resistivity was measured by the standard four-terminal method. Before each measurement of the physical properties, the surface of the single crystals was polished to remove the Bi flux and to obtain a clean surface, which tends to be degraded by air or moisture. Resistivity measurements under low temperature ($T < 1.8$ K) and magnetic field were performed using a custom-developed dilution refrigerator and a superconducting magnet (Oxford Instruments). The resistivity was measured by the standard four-terminal method using an LR-700 (Linear Research) or Model 372 (Lake Shore Cryotronics, Inc.).

**Specific heat capacity measurements**

Specific heat measurements at low temperature were performed by the thermal relaxation method using a home-made calorimetry cell. The cell was tightened on the mixing chamber of the dilution refrigerator. A calibrated thermometer (2 k$\Omega$-RuO$_2$ chip resistor, KOA Corporation) and a heater (120 $\Omega$-strain gauge, Kyowa Electronic Instruments Co., Ltd) were attached to the sample stage with stycast 1266. The sample stage (2×2.5×0.1 mm$^3$ Ag plate) was suspended in the vacuum space by manganin wires, which serve both as thermal leak paths to the bath and current leads. The temperature dependence of the addenda heat capacity and the thermal conductivity of the manganin were determined beforehand by a measurement without sample. The SrAuBi sample was attached on the stage with Apiezon N grease. We measured several relaxation processes covering the temperature range of 0.7 K $< T <$ 3 K, and obtained the sample + addenda heat capacity $C_{s+a}$ from the relationship

$$C_{s+a} = -\frac{A}{l}\int_{Tbath}^{T} K(T')dT / (\frac{dT(t)}{dt}).$$

Here, the cross-section ($A$) and the length ($l$) of the leak path, the temperature of the bath ($T_{bath}$), and the thermal conductivity of the manganin ($\kappa$) are all known parameters, and the time derivative of the stage temperature $dT(t)/dt$ is obtained from the relaxation curve. We finally obtained the sample heat capacity by subtracting the addenda heat capacity from $C_{s+a}$.

**Band calculations and Structure optimization**

The relativistic and non-relativistic bulk electronic structure calculations and relaxed structural of SrAuBi were performed within the density functional theory (DFT) using the Perdew-Burke-Ernzerhof (PBE) exchange-correlation functional as implemented in the Quantum ESPRESSO code[64–67]. The projector augmented wave (PAW) method has been used to account for the treatment of core electrons[67]. The cut-off energy for plane waves forming the basis set was set to 50 Ry. The lattice parameters and atomic positions were taken from the single-crystal XRD experiment. To consider the strong on-site Coulomb interaction of the Au-5$d$ states, an effective Hubbard-like potential term $U_{eff}$ was added. The value of $U_{eff}$ was fixed at 4 eV for the Au-5$d$ orbitals[68]. The corresponding Brillouin zone was sampled using a $12 \times 12 \times 9$ $k$-mesh to calculate the PWscf, and a $20\times20\times20$ $k$-mesh to perform Fermi surface visualization by the FermiSurfer package[69].

**ACKNOWLEDGEMENTS**

The authors thank I. Terasaki, S. Fujimoto, T. Mizushima, K. Aoyama, and K. Akiba for fruitful discussions. Single-crystal XRD measurements were conducted with the approval of the Japan Synchrotron Radiation Research Institute (JASRI) (Proposal No. 2021B1198 and 2022A1158). The authors would like to thank M. Arai, Y. Minowa, and M. Asida for their assistance in the SEM/EDX measurements. This study was supported in part by JSPS KAKENHI (Grant No. JP20K03802, JP21H01030, JP22H00343, JP23H04871, and JP23H04868), the Asahi Glass Foundation, Research Foundation for the Electrotechnology of Chubu, and the Murata Science Foundation.

**AUTHOR CONTRIBUTIONS**

H.T. conceived and led the project. H.T., T.S., and M.T. synthesized the samples and performed magnetic and transport mesurements. A.N. measured single-crystal XRD measurements. K.A. and T.C.K measured specific heat capacity and low temperature transport properties below 1.8 K. H.T. measured the SEM/EDX. H. T., T.S., A.H.M, and M.O. conducted the band calculations and structural optimization. S.I. supervised the work. H.T. and S.I. prepared the manuscript with notable inputs from all authors.

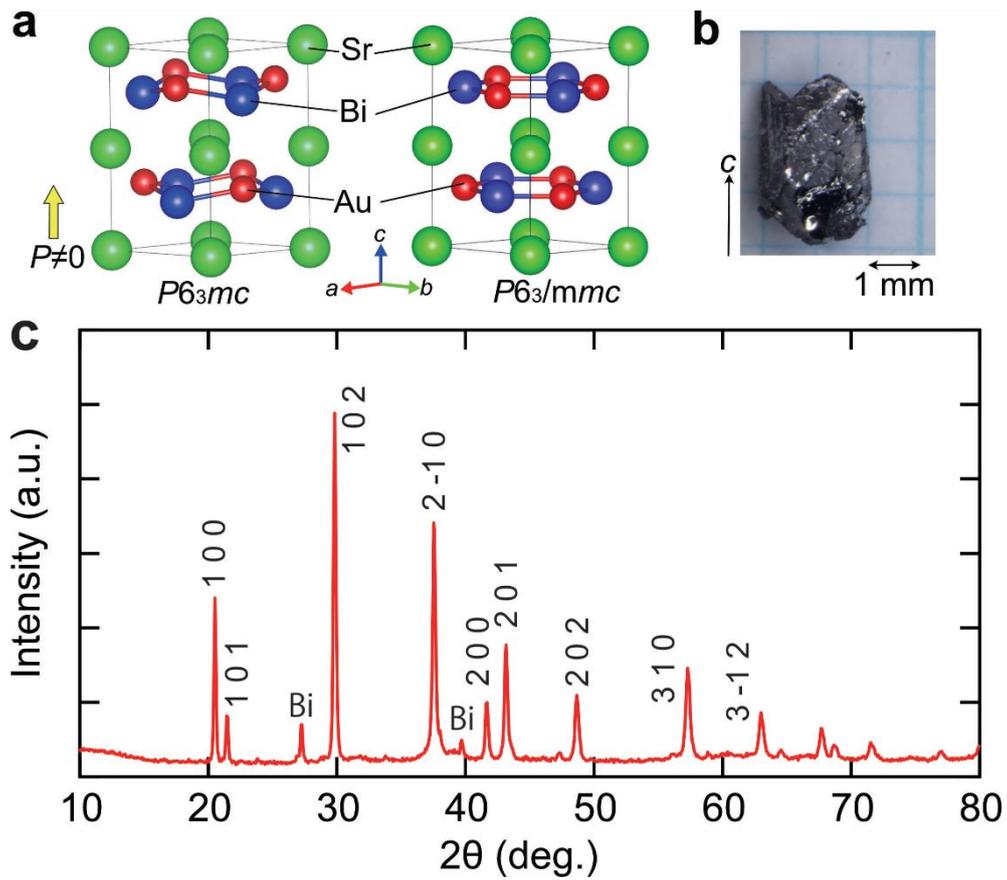

**Fig. 1: a** Crystal structures of a low-temperature polar ($P6_3mc$) phase and a high-temperature nonpolar ($P6_3/mmc$) phase of SrAuBi. **b** The single crystal of SrAuBi. **c** Powder x-ray diffraction pattern of SrAuBi at room temperature.

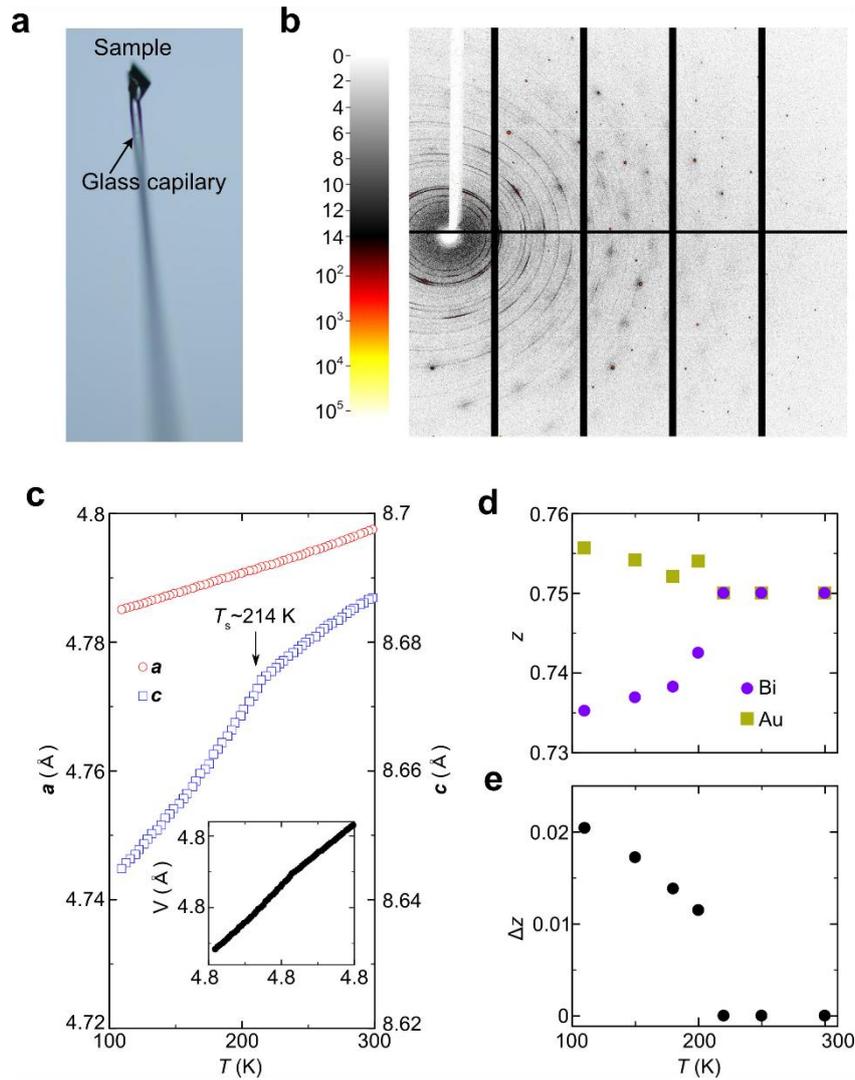

**Fig. 2**: **a** Photo of a single crystal for the X-ray diffraction measurement. **b** Single crystal X-ray diffraction pattern. **c** Temperature dependence of the lattice parameters. Temperature dependence of (**d**) the site position of $z$ for Bi and Au and (**e**) the difference of $z$ ($\Delta z$) between those of Bi and Au.

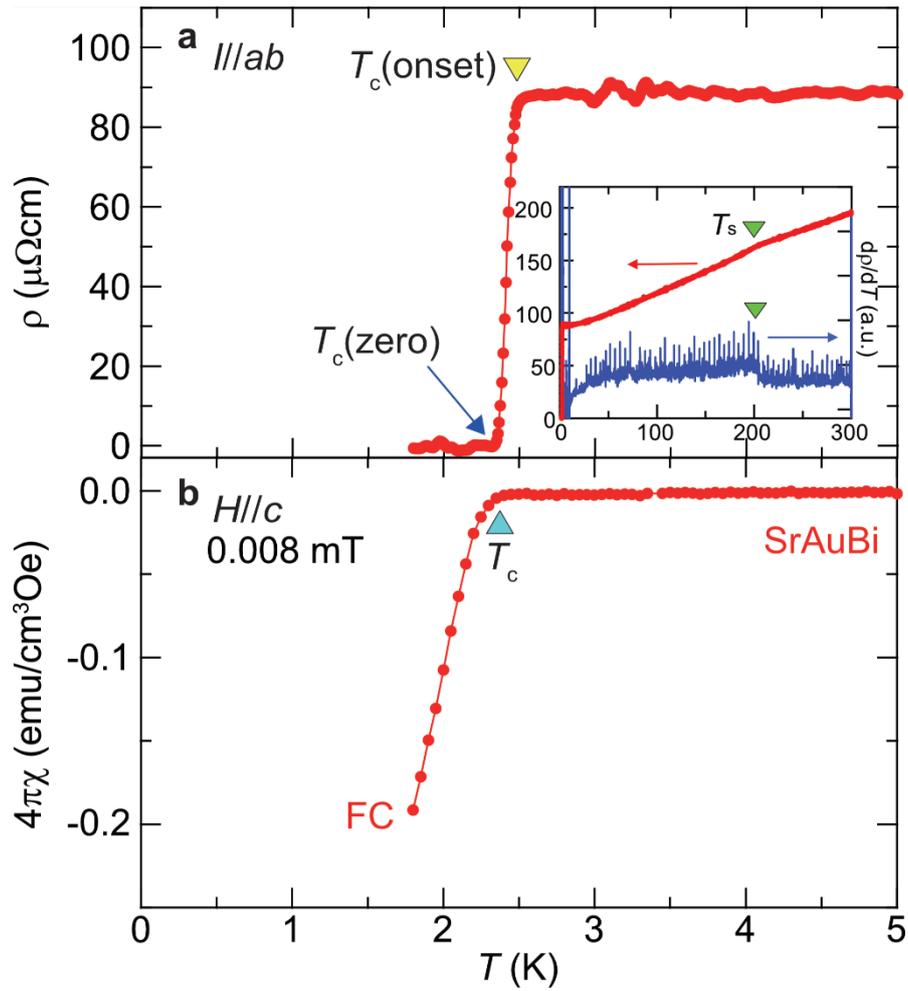

**Fig. 3**: **a** Temperature dependence of in-plane resistivity below 5 K. Inset shows the temperature dependence of in-plane resistivity (ρ) and temperature derivative for the resistivity (dρ/dT) from 300 to 1.8 K. **b** Magnetic susceptibility under H//c at 0.008 mT below 5 K.

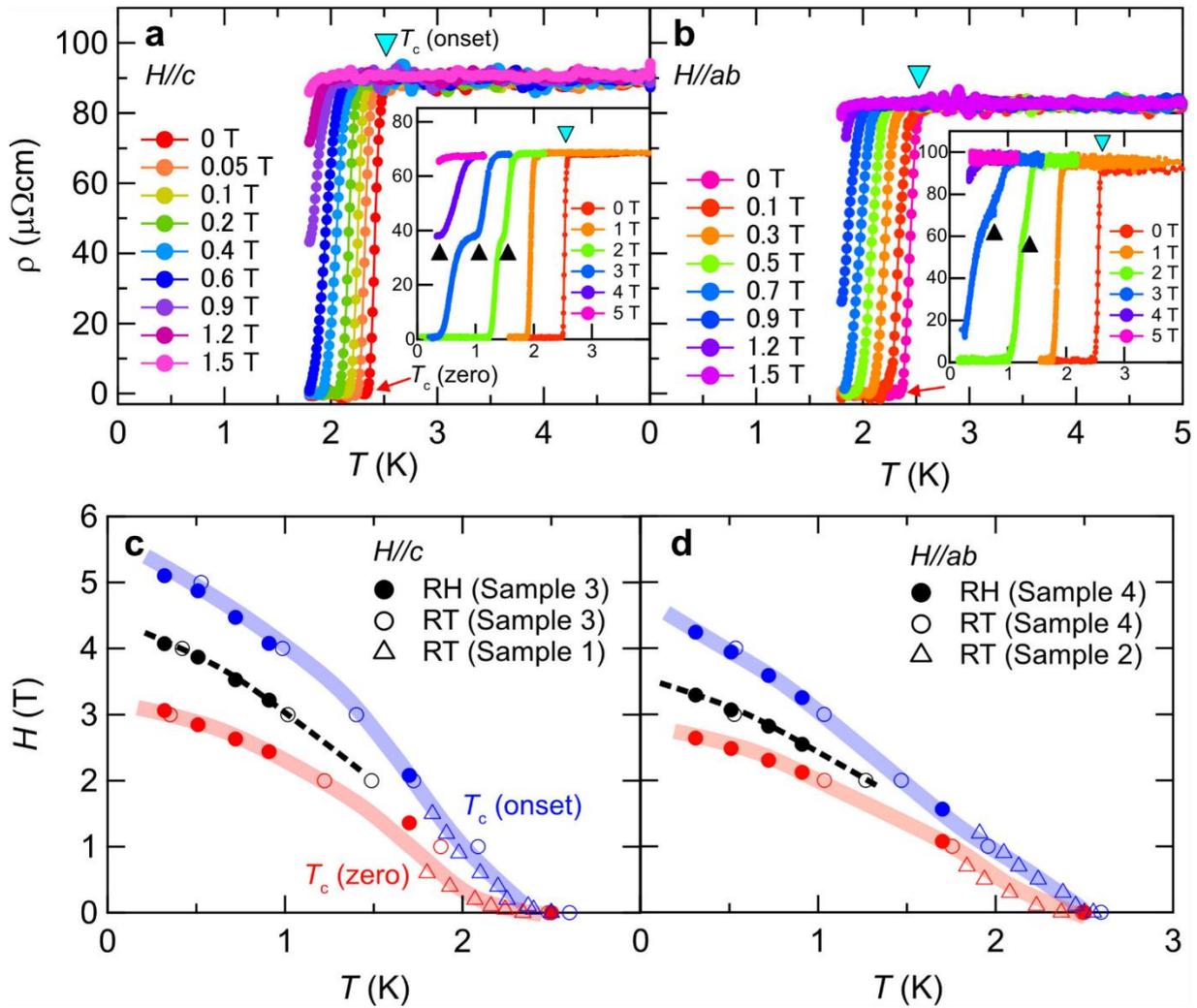

**Fig. 4:** Temperature dependence of in-plane resistivity below 5 K at various magnetic fields for (**a**) $H//c$ and (**b**) $H//ab$ with different samples No. 1 and No. 2, respectively. Insets of (**a**) and (**b**) show the in-plane resistivity down to 0.3 K for $H//c$ (sample No. 3) and $H//ab$ (sample No. 4), respectively. Temperature dependence of the upper critical field $H_c$ is defined by the zero resistivity and onset of superconducting transition (**c**) $H//c$ and (**d**) $H//ab$. The data of RH is shown in the Supplementary Fig. 9.

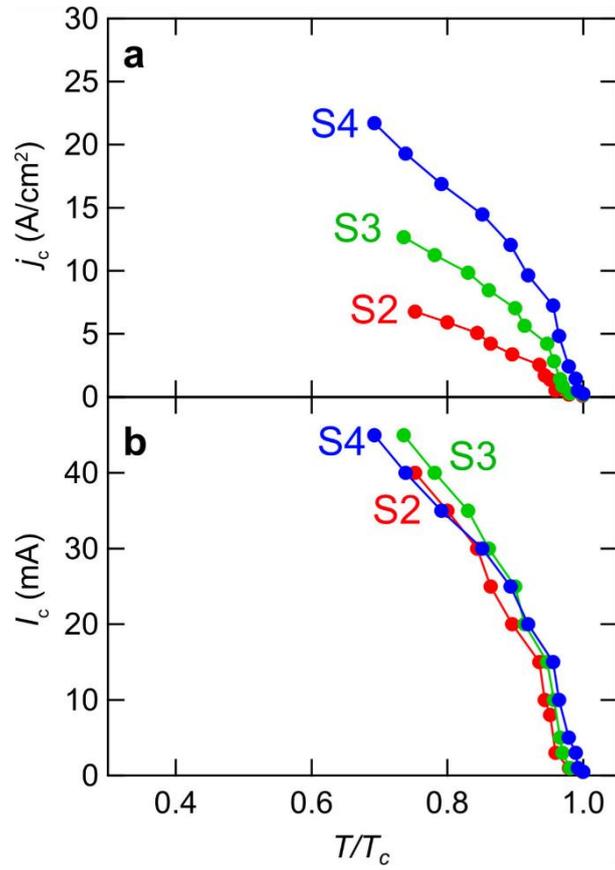

**Fig. 5: a** The temperature-dependence of critical current density of one sample with different thicknesses. **b** The temperature-dependence of critical current of one sample with different thicknesses.

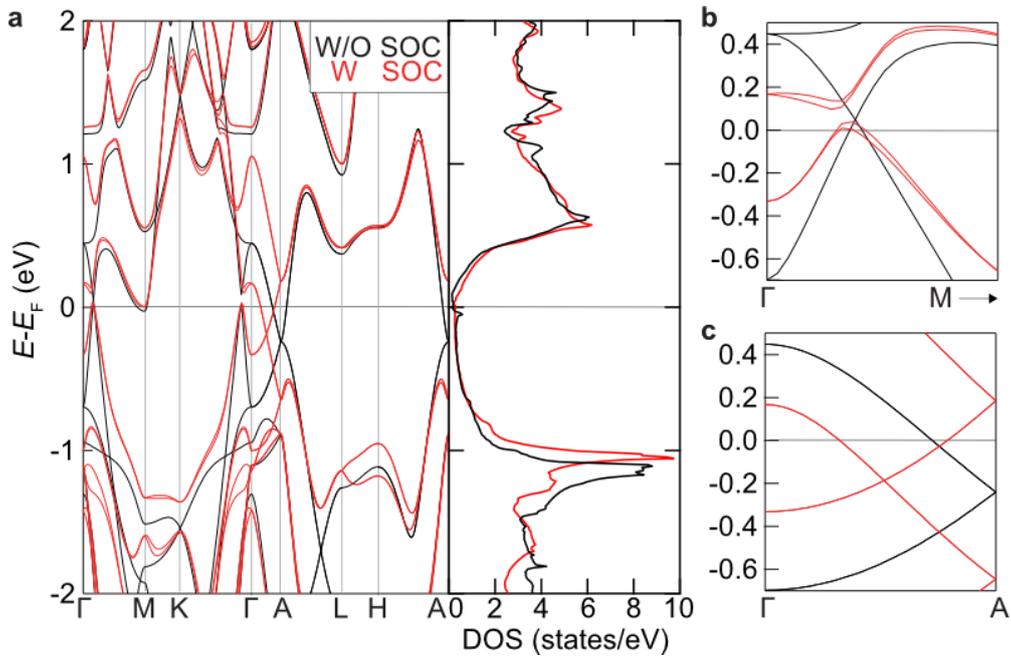

**Fig. 6: a** The electronic structure and density of states (DOS) with and without spin-orbit coupling (SOC) of SrAuBi. Magnified view of the bands near the Fermi level along (**b**) Γ–M and (**c**) Γ–A lines.